\documentclass[prl,preprint,graphicx,citesort,amssymb]{revtex4}
\usepackage{amssymb}
\usepackage{graphicx}
\usepackage{verbatim}
\usepackage{amstext}
\RequirePackage{ifthen}
\newcommand{\ite}{\ifthenelse}

\begin{document}
\title{Crossover of high and low spin states in transition metal complexes}

\author{Hannes Raebiger} 
\affiliation{Department of Physics, Yokohama National University, Yokohama 240-8501, Japan}
\author{Shuhei Fukutomi} 
\affiliation{Department of Physics, Yokohama National University, Yokohama 240-8501, Japan}
\author{Hiroshi Yasuhara} 
\affiliation{Institute of Materials Research, Tohoku University, Sendai 980-8577, Japan}

\begin{abstract}
The stability of high vs.\ low spin states of transition metal complexes 
has been interpreted by ligand field theory,
which is a perturbation theory
of the electron-electron interaction. 
The present first principles calculation
of a series of five cobalt complexes 
shows that
the electron-electron interaction energy difference between the two states
(i) exhibits the opposite trend to the total energy difference as the ligand nuclear charge varies,
and (ii) is three or four orders of magnitude greater than the total energy difference.
A new interpretation
of the crossover of high and low spin states is given in terms
of the chemical bonding.
\end{abstract}
\maketitle

Transition metal complexes such as the octahedral CoL$_6$ exhibit both 
high and low spin states. 
The energy difference between these two states
$\Delta E = E^{\rm HS} - E^{\rm LS}$ 
depends on the atomic number $Z_L$ of ligand L;
the high spin state of $E^{\rm HS}$ is stable for large $Z_L$ and low spin state of $E^{\rm LS}$ for small $Z_L$.
For almost a century, 
this crossover has been interpreted by ligand field theory
which is a perturbation theory
ascribing
$\Delta E$ to the electron--electron repulsion energy difference $\Delta V_{ee} = V_{ee}^{\rm HS} - V_{ee}^{\rm LS}$
between the two states~\cite{%
Bethe:1929p2924,SLATER:1929p2076,VanVleck:1935p807,Tanabe:1954p2794,Griffith:1957p381}.
Our quantum chemical calculation of a series of CoL$_6$ complexes, however, shows that,
ligand field theory and any other attempts to ascribe $\Delta E$ to $\Delta V_{ee}$ are invalid,
because
$\Delta V_{ee}$ (i) never changes sign,
(ii) decreases with decreasing $Z_L$ 
and thus exhibits the opposite trend to $\Delta E$,
and 
(iii) is three or four orders of magnitude greater than $\Delta E$, 
clearly beyond perturbation theory.
Correctly,
the spin multiplicity is variationally determined by an intricate 
interplay
between the electron--electron repulsion $V_{ee}$, 
the electro-nuclear attraction $V_{ne}$, and the nucleus--nucleus repulsion $V_{nn}$.
In conclusion,
the crossover of high and low spin states is a consequence of different Co--L bondings, ionic or covalent,
which is found by an accurate treatment of
Coulomb correlation between ligand $p$ and cobalt $d$ electrons
in the present calculation.

Quantum chemical first principles calculations were carried out using the GAMESS~\cite{Schmidt:1993p2889} package.
Coulomb correlation is included via the 
complete active space self-consistent field (CAS-SCF) method,
where full configuration interaction calculations are carried out within
the chosen active spaces.
Our active spaces are chosen from the transition atom 3$d$, 4$d$ and ligand 2$p$ orbitals [CAS(12,10)]; 
in CAS(10,6) the ligand 2$p$ orbitals are excluded.
For each complex and spin state, the molecular geometry is fully optimized.
All calculations satisfy the virial theorem $2T+V = 0$ 
with the virial ratio $V/T = -2.000 \pm 0.001$.
For F in [CoF$_6$]$^{3-}$ and N in [Co(CN)$_6$]$^{3-}$ we use the basis set 6-31G$^{++**}$,
and 6-31G$^{**}$ for all other atoms.

We study five octahedral cobalt complexes:
[CoF$_6$]$^{3-}$, 
[Co(OH$_2$)$_6$]$^{3+}$,
[Co(NH$_3$)$_6$]$^{3+}$,
[Co(CN)$_6$]$^{3-}$,
and [Co(CO)$_6$]$^{3+}$.
The first two
have quintet ($S=2$; high spin) ground states, and the remaining three 
singlet ($S=0$; low spin) ground states.~\cite{Pauling:1931p1367,Pauling:1948p191,Pauling:1960}
The calculated energy differences $\Delta E = E^5 - E^1$ between the quintet and singlet 
states
for the five complexes
are given in table~\ref{tab:delta}.
Obviously,
the singlet ground states of 
[Co(NH$_3$)$_6$]$^{3+}$,
[Co(CN)$_6$]$^{3-}$,
and [Co(CO)$_6$]$^{3+}$ 
are predicted correctly
only if Coulomb correlation is 
considered for  
those ligand L $p$ electrons 
which form $\sigma$ bonds with the central cobalt atom
as well as for the Co $d$ electrons [CAS(12,10)];
the inclusion of Coulomb correlation only for the Co $d$ electrons [CAS(10,6)]
leads to wrong ground states.

\begin{table}[t]
\caption{Energy difference $\Delta E$ between high and low-spin states in Hartree atomic units,
evaluated in Hartree-Fock (HF) 
and complete active space self-consistent field [CAS(12,10) and CAS(10,6)] calculations.}
\begin{tabular}{l c c c}
\hline \hline
Complex & HF & CAS(10,6) & CAS(12,10) \\
\hline
${\rm[CoF_6]^{3-}}$
& $-$0.11 & $-$0.08 & $-$0.05	\\
${\rm [Co(OH_2)_6]^{3+}}$ 
& $-$0.09 & $-$0.06	& $-$0.03 \\
${\rm [Co(NH_3)_6]^{3+}}$ 
& $-$0.06 & $-$0.03 & 0.01 \\
${\rm [Co(CN)_6]^{3-}}$ 
& $-$0.04 & $-$0.01 & 0.05 \\
${\rm [Co(CO)_6]^{3+}}$ 
& $-$0.06 & $-$0.03	& 0.02 \\
\hline \hline
\end{tabular}
\label{tab:delta}
\end{table}

The energy differences $\Delta E$ (table~\ref{tab:delta})
are as small as some tens of milliHartree,
while the total energies are as large as 2000 Hartree per complex.
Earlier studies of high and low spin states have relied on perturbation theory~\cite{Heitler:1927p455,
Bethe:1929p2924,SLATER:1929p2076,VanVleck:1935p807,Tanabe:1954p2794,
Heisenberg:1928p2886,Dirac:1929p2885,VanVleck:1934p2617,Griffith:1957p381}
instead of calculating $\Delta E$ from the total energies
$E^{2S+1} = T^{2S+1} + V_{ne}^{2S+1} + V_{nn}^{2S+1} + V_{ee}^{2S+1}$
for each $S$.
These earlier studies are based on the assumption 
that
$T^{2S+1} + V_{ne}^{2S+1} + V_{nn}^{2S+1}$ 
has no dependence on the spin multiplicity $2S+1$,
and that $\Delta E$ can be ascribed solely to $V_{ee}^{2S+1}$, i.e. $\Delta E \approx \Delta V_{ee}$.
In this work, the potential energy difference and its components 
$\Delta V$ ($ = \Delta V_{ee} + \Delta V_{ne} + \Delta V_{nn}$)
given in table~\ref{tab:delta2}
are calculated
under the virial theorem condition
$E^{2S+1} = \frac{1}{2} V^{2S+1} = \frac{1}{2} (V_{ne}^{2S+1} + V_{nn}^{2S+1} +  V_{ee}^{2S+1})$,
satisfying the virial ratio $V^{2S+1}/T^{2S+1} = -2.000\pm0.001$ up to at least three digits for both 
the high and low spin states.
We find that
the leading contribution to $\Delta E$ is not $\Delta V_{ee}$ but $\Delta V_{ne}$ 
that comes from the difference in the electron density distribution between the two states.
$\Delta V_{ne}$ is the only energy difference component that exhibits
an increasing trend with decreasing $Z_L$ in a similar way as $\Delta E$;
$\Delta V_{ee}$ and $\Delta V_{nn}$ exhibit the opposite trend.
However,
none of $\Delta V_{ee}$,  $\Delta V_{nn}$, and $\Delta V_{ne}$ individually changes sign
with decreasing $Z_L$,
and each of them is three or four orders larger than $\Delta E$ in magnitude.
Thus, the crossover of high and low spin states 
is the outcome of a delicate interplay of all three
potential energy components.

\begin{table}
\caption{Potential contributions $\Delta V_{ee}$, $\Delta V_{nn}$, and $\Delta V_{ne}$ to 
the energy difference $\Delta E$ in Hartree atomic units
evaluated in HF and CAS(12,10) calculations.}
\begin{tabular}{l l c c c}
\hline \hline
Complex &  & $\Delta V_{ee}$ & $\Delta V_{nn}$ & $\Delta V_{ne}$\\ 
\hline
${\rm[CoF_6]^{3-}}$
&	HF&		$-$27.47&		$-$24.37&	51.48 \\
&	CAS(12,10)&	$-$27.48&		$-$23.5&	50.84 \\
${\rm [Co(OH_2)_6]^{3+}}$
&	HF&		$-$31.76&		$-$30.54&	62.01 \\
&	CAS(12,10)&	$-$33.31&		$-$31.23&	64.31 \\
${\rm [Co(NH_3)_6]^{3+}}$ 
&	HF&		$-$37.02&		$-$35.08&	71.95 \\
&	CAS(12,10)&	$-$38.81&		$-$36.29&	75.21 \\
${\rm [Co(CN)_6]^{3-}}$ 
&	HF&		$-$54.42&		$-$49.72&	103.95 \\
&	CAS(12,10)&	$-$69.27&		$-$64.19&	133.51 \\
${\rm [Co(CO)_6]^{3+}}$  
&	HF&		$-$58.58&		$-$57.59&	115.98 \\
&	CAS(12,10)&	$-$72.34&		$-$71.69&	144.07 \\
\hline \hline
\end{tabular}
\label{tab:delta2}
\end{table}

In order to clarify why the ground state varies
from high to low spin states,
we give a detailed analysis of the potential energy $V$ and its components $V_{ee}$, $V_{nn}$, and $V_{ne}$.
We calculate $E^{2S+1}$
by both Hartree-Fock (HF)
and complete active space self-consistent field (CAS-SCF) methods.
Electrons tend to avoid each other due to Pauli's exclusion principle (Fermi correlation)
and due to Coulomb repulsion (Coulomb correlation).
Fermi correlation is already accounted for in HF 
and is strongest when the number of spin-parallel electrons is largest.
On the other hand, Coulomb correlation is strongest when the number of spin-parallel electrons is smallest
and hence causes the crossover of high and low spin states.
Coulomb correlation not only reduces the short-range interelectronic contribution of $V_{ee}$,
but also affects $V_{ne}$ and $V_{nn}$.~\cite{RussellJBoyd:1984p1102,Katriel:1977p2888,Maruyama:2008p2144,Oyamada:2010p2874}
Coulomb correlation gives rise to the Coulomb hole between spin-antiparallel electrons
and at the same time deepens the Fermi hole between spin-parallel electrons.
These correlation holes have an effect to reduce the Hartree-Fock screening of the nuclei at short interelectronic distances,
leading to a contraction of the electron density distribution around individual nuclei 
as well as to a change in the equilibrium nuclear configuration.
The correlation effects on $V_{ee}$, $V_{nn}$, and $V_{ne}$ are described in the following.

The potential energy difference $\Delta V_{nn}$ between high and low spin states
arises from a change in the equilibrium nuclear configuration $\{\vec R_I\}$.
As is seen from table~\ref{tab:geom}, the bond lengths between the central cobalt atom and the six ligands 
are always larger for the quintet state and hence $V_{nn}^5 < V_{nn}^1$.
It means that the high spin state complexes are larger in size than the low spin state ones.
Therefore, the average value of the
electron--electron separation $|\vec r_i - \vec r_j|$ is enlarged for the high spin state complexes,
i.e.\ $V_{ee}^5 < V_{ee}^1$.
Similarly, the average value of the electron--nucleus distances $|\vec R_I - \vec r_j|$ tends to be reduced 
in the low spin state complexes,
i.e.\ $V_{ne}^5 > V_{ne}^1$.
This trend is observed for both HF and CAS-SCF and explains why
the repulsive $V_{ee}$ and $V_{nn}$ favor high spin states 
and the electron--nucleus attraction low spin states.
Next we examine which effect
causes the change in sign of $\Delta E$,
i.e. the crossover of high and low spin states for $Z_L < 8$.

\begin{table}
\caption{Equilibrium Co--L bond lengths in Bohr atomic units for high and low spin states (HS and LS)
evaluated in HF and CAS(12,10) calculations.
Where two values are given, the $z$ bond is shorter and the $x$ and $y$ bonds are equally long.
Where three values are given, the $x$, $y$ and $z$ bonds are inequivalent.}
\begin{tabular}{llll}
\hline \hline
Complex &  & HF & CAS(12,10)  \\ 
\hline
${\rm[CoF_6]^{3-}}$ & ÊÊÊÊÊÊÊLS ÊÊÊÊÊÊÊÊÊÊÊ& 3.592 & 3.592 ÊÊÊÊ\\
ÊÊÊÊÊÊÊÊÊÊÊÊÊÊÊÊÊÊÊÊÊÊÊÊÊÊÊÊÊÊÊÊÊ& ÊÊÊÊÊÊÊHS ÊÊÊÊÊÊÊÊÊÊÊ& 3.623, 3.795
& 3.621, 3.789 ÊÊÊ\\
${\rm [Co(OH_2)_6]^{3+}}$ & ÊÊÊÊÊÊÊÊÊÊÊÊÊÊLS ÊÊÊÊÊ& 3.634 & 3.621 ÊÊÊÊÊÊÊÊÊÊÊ
ÊÊÊÊÊÊÊ\\
ÊÊÊÊÊÊÊÊÊÊÊÊÊÊÊÊÊÊÊÊÊÊÊÊÊÊÊÊÊÊÊÊÊ& ÊÊÊÊÊÊÊHS ÊÊÊÊÊÊÊÊÊÊÊ& 3.812 & 3.800
ÊÊÊ\\
${\rm [Co(NH_3)_6]^{3+}}$ Ê& ÊÊÊÊÊÊÊLS ÊÊÊÊÊÊÊÊÊÊ& 3.853 & 3.838 ÊÊÊÊÊÊÊÊÊÊÊ
ÊÊÊÊÊÊÊÊÊÊ\\
ÊÊÊÊÊÊÊÊÊÊÊÊÊÊÊÊÊÊÊÊÊÊÊÊÊÊÊÊÊÊÊÊÊ& ÊÊÊÊÊÊÊHS ÊÊÊÊÊÊÊÊÊÊÊ& 4.052, 4.116
& 4.078, 4.086, 4.097 ÊÊÊÊ\\
${\rm [Co(CN)_6]^{3-}}$  ÊÊ& ÊÊÊÊÊÊÊLS ÊÊÊÊÊÊÊÊÊ& 3.878 & Ê3.766 ÊÊÊÊÊÊÊÊÊÊÊ
ÊÊÊÊÊÊÊ\\
ÊÊÊÊÊÊÊÊÊÊÊÊÊÊÊÊÊÊÊÊÊÊÊÊÊÊÊÊÊÊÊÊÊ& ÊÊÊÊÊÊÊHS ÊÊÊÊÊÊÊÊÊÊÊ& 4.144, 4.225
& 4.137, 4.188 ÊÊÊ\\
${\rm [Co(CO)_6]^{3+}}$   ÊÊÊ& LS ÊÊÊÊÊÊÊÊÊÊÊÊÊÊÊ& 3.946 & Ê3.840 ÊÊÊÊÊÊÊÊÊÊÊ
ÊÊÊÊÊÊÊÊÊÊÊ\\
ÊÊÊÊÊÊÊÊÊÊÊÊÊÊÊÊÊÊÊÊÊÊÊÊÊÊÊÊÊÊÊÊÊ& ÊÊÊÊÊÊÊHS ÊÊÊÊÊÊÊÊÊÊÊ& 4.269, 4.309
& 4.263 ÊÊÊ\\
\hline\hline
\end{tabular}
\label{tab:geom}
\end{table}

\begin{figure} %
   \centering
   \includegraphics{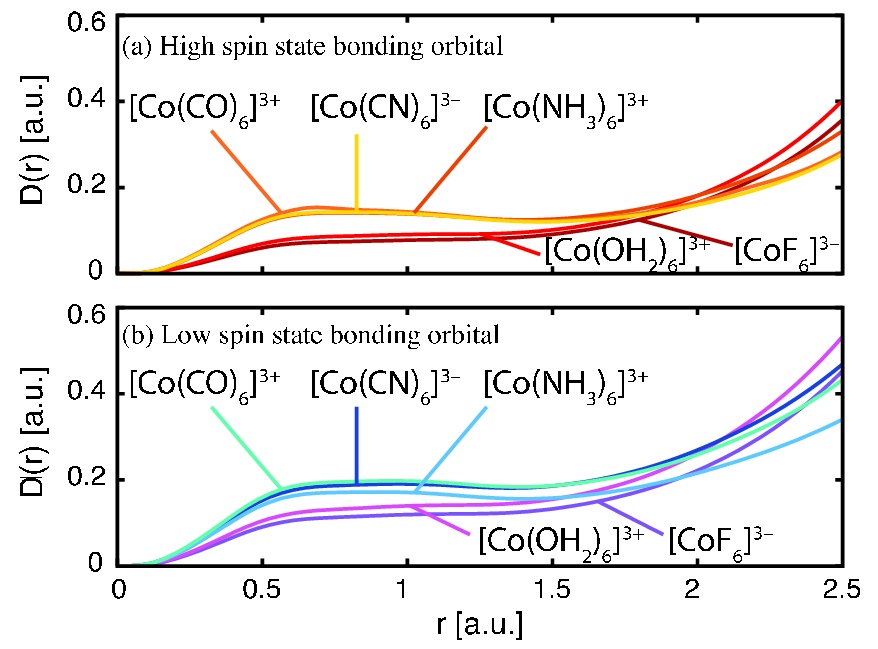} 
   \caption{Charge density distribution $D(r)$ of bonding orbitals for each complex in Hartree atomic units
   obtained in HF calcuation.
   (a) High spin states. (b) Low spin states.}
   \label{fig:sigma}
\end{figure}

For complexes with $Z_L \geq 8$,
the sum of $\Delta V_{nn}$ and $\Delta V_{ee}$ is greater than $\Delta V_{ne}$
because Fermi correlation that tends to maximize the spin multiplicity~\cite{Katriel:1977p2888} dominates
over Coulomb correlation. 
As $Z_L$ becomes smaller than 8, the Coulomb correlation effects become larger and larger, 
especially for $\Delta V_{ne}$, which overwhelms the sum of $\Delta V_{nn}$ and $\Delta V_{ee}$.
We find that this remarkable increase in $\Delta V_{ne}$ is due to a change in the nature of the chemical bond:
the bonding $\sigma$ orbitals are ionic
for $Z_L \geq 8$ and covalent $Z_L < 8$.
In Fig.~1 we show the radial distribution $D(r)$ of these $\sigma$ orbitals.
There is a striking difference between the
complexes 
with $Z_L \geq 8$ and $Z_L < 8$.
$D(r)$ for the complexes with $Z_L < 8$ exhibits a maximum close to the central cobalt,
followed by a minimum around the bond center, and an increase towards another maximum
close to the ligand. 
This two-hump shape is a typical example of the charge density distribution in the covalent bond.
On the other hand,
$D(r)$ of the fluorine and water complexes
only has one maximum close to the ligand, and monotonously decreases towards the cobalt atom,
typical of the ionic bond.
This terminology of ionic and covalent bonds 
is consistent with Pauling's description of the very same complexes,~\cite{Pauling:1948p191,Pauling:1960}
but the fact that covalent complexes tend to exhibit low spin states and ionic complexes high spin states
still remains undescribed.
We unveil this mechanism in the following:

\begin{figure} 
   \centering
   \includegraphics[width=3.25in]{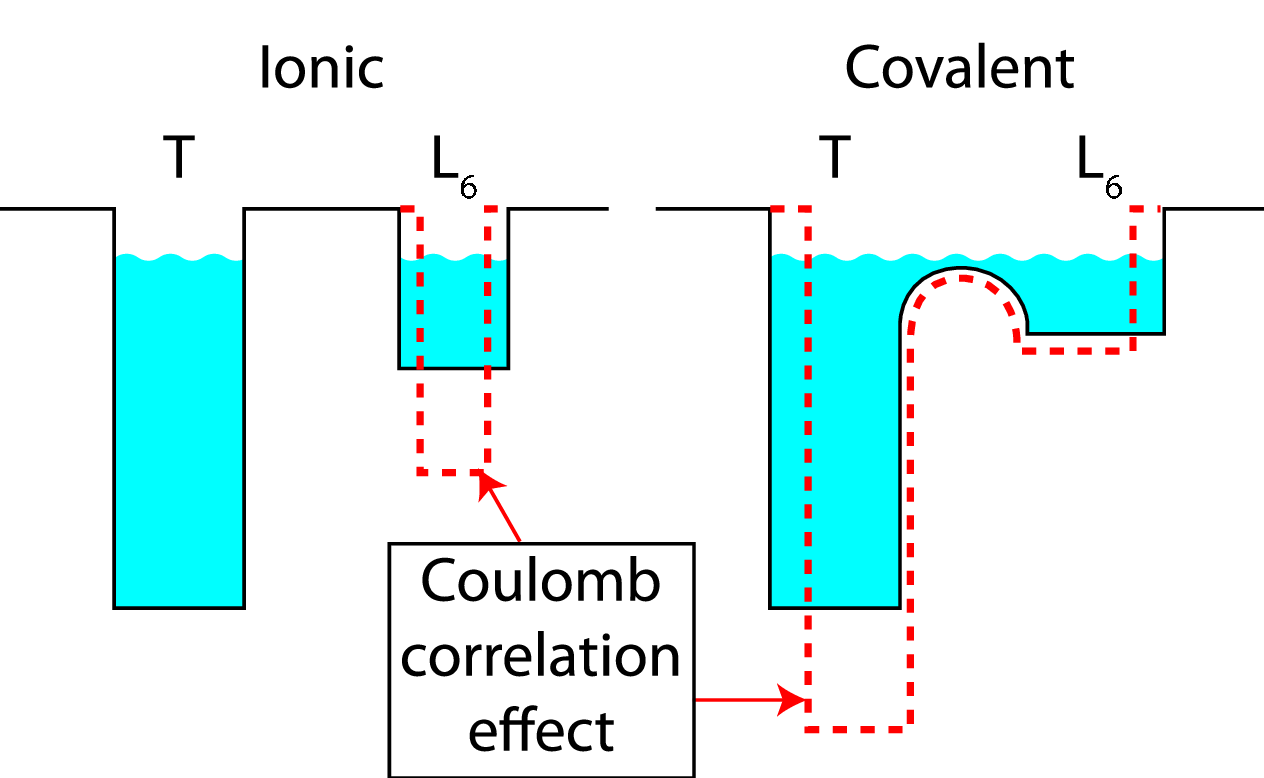} 
   \caption{Potential box illustration of ionic and covalent complexes.
   Both the transition metal T and ligand L$_6$ potential boxes are filled up to the same electron chemical potential.
   The solid black line illustrates the HF calculation, 
   and the red dashed line illustrates the deepening of the potential boxes due to Coulomb correlation included via CAS(12,10).
   The internuclear distance between T and L$_6$ is widened for Ionic and reduced for Covalent by the action of correlation.}
   \label{fig:boxes}
\end{figure}

Let us illustrate the central transition atom T and the octahedral ligand L$_6$ 
as two potential boxes (Fig.~2).
The covalent complex is depicted as a set of two connected boxes,
and the ionic complex as two separated boxes, which are filled up to the same electron chemical potential.
This illustration represents the qualitative differences in the nature of the $\sigma$ orbitals (Fig.~1),
which yield qualitatively
different Coulomb correlation effects.
Coulomb correlation enhances electron localization, which is depicted as a deepening of the boxes.
For the covalent complex,
the T and L$_6$ boxes are both deepened, while for the ionic complex, only the L$_6$ box is deepened,
corresponding to whether the $\sigma$ orbital distribution (Fig.~1) has two maxima or one.
Since for the ionic complex the main Coulomb correlation effect occurs in the ligands,
the central transition atom essentially behaves as an isolated atom or ion, where 
the highest spin multiplicity state has the lowest energy due to Fermi correlation.~\cite{Katriel:1977p2888}
On the other hand, for the covalent complex, 
also the T box becomes deeper and moreover this deepening is larger at T than at L$_6$.
This is because the attractive potential due to nucleus $I$,
$V_{ne,I} = - \sum_j Z_I/|\vec R_I - \vec r_j|$,
is proportional to the nuclear charge $Z_I$,
and hence
an enhancement in electron localization at T yields a larger energy 
gain than it would at L$_6$, because $Z_T > Z_L$. 
Therefore, only for the covalent complex, the electronic configuration of the central transition atom T
is strongly influenced by the spin-paired $\sigma$ bonding orbitals formed mostly of ligand $p$ electrons
and hence the low spin state is the ground state.

Our calculations show that octahedral cobalt ($Z_{\rm Co}=27$) complexes with
ligands of $Z_{\rm L}<8$ exhibit low spin states in accordance with experiment.
The energy gain that stabilizes the low spin states of these cobalt complexes arises from 
Coulomb correlation effects in 
the covalent Co--L bonding that lowers
$V_{ne,{\rm Co}} = - \sum_j Z_{\rm Co}/|\vec R_{\rm Co} - \vec r_j|$.
Obviously, this energy gain is expected to be smaller for lighter transition atoms because of the proportionality $V_{ne,I} \propto Z_I$.
Indeed, octahedral complexes of iron ($Z_{\rm Fe}=26$) with the ligand $Z_{\rm L}<7$ exhibit low spin states.
The crossover of high and low spin states is accompanied with a change from ionic to covalent bonding.
We have demonstrated that
the electron density distribution and the equilibrium nuclear configuration are strikingly different between the ionic and covalent bondings.
These differences are accompanied with changes in $V_{ne}$ and $V_{nn}$,
which are as significant as changes in $V_{ee}$.
To conclude,
theories relying on
$V_{ee}$ alone predict wrong ground states,
so earlier works~\cite{Bethe:1929p2924,SLATER:1929p2076,VanVleck:1935p807,Dirac:1929p2885,
Heitler:1927p455,Heisenberg:1928p2886,VanVleck:1934p2617,Tanabe:1954p2794,Griffith:1957p381}
together with any textbook description of high and low spin states
neglecting $V_{ne}$ and $V_{nn}$ are invalid. 

{\em Acknowledgements.}
We thank M.~Tachikawa, Y.~Kita, S.~Ishii and K.~Shudo for inspiring discussions, 
M.W.~Schmidt for technical support,
and Y.~Kawazoe for support.
This work was funded by a Grant-in-Aid for Young Scientists (A) grant (No. 21686003) 
from the Japan Society for the Promotion of Science.

\end{document}